# MgB$_2$ Energy Gap Determination by Scanning Tunneling Spectroscopy

T.W. Heitmann[1,2], S.D. Bu[3], D.M. Kim[3], J.H. Choi[3], J. Giencke[3], C.B. Eom[1,2,3], K.A. Regan[4], N. Rogado[4], M.A. Hayward[4], T. He[4], J.S. Slusky[4], P. Khalifah[4], M. Haas[4], R.J. Cava[4], D.C Larbalestier[2,3], M.S. Rzchowski[1,2]

[1]*Physics Department, University of Wisconsin*
*1150 University Ave., Madison, WI  53706*
[2]*Applied Superconductivity Center , University of Wisconsin*
*1500 Engineering Dr., Madison, WI  53706*
[3]*Dept. of Materials Science and Engineering, University of Wisconsin*
*1509 University Ave., Madison, WI  53706*
[4]*Department of Chemistry and Princeton Materials Institute, Princeton University*
*70 Prospect Ave., Princeton, NJ  08540*

We report scanning tunneling spectroscopy (STS) measurements of the gap properties of both ceramic MgB$_2$ and *c*-axis oriented epitaxial MgB$_2$ thin films. Both show a temperature dependent zero bias conductance peak and evidence for two superconducting gaps. We report tunneling spectroscopy of superconductor-insulator-superconductor (S-I-S) junctions formed in two ways in addition to normal metal-insulator-superconductor (N-I-S) junctions. We find a gap $\Delta$=2.2-2.8 meV, with spectral features and temperature dependence that are consistent between S-I-S junction types. In addition, we observe evidence of a second, larger gap, $\Delta$=7.2 meV, consistent with a proposed two-band model.

I. Introduction

Since its discovery as a superconductor with $T_C$=39 K by Akimitsu et al[1], magnesium diboride has garnished considerable interest. Unlike most known superconductors, $MgB_2$ is a metallic binary compound with a relatively simple layered hexagonal crystal structure. Its relatively high $T_C$ coupled with its simple structure (hence, ease of processing) and high critical current density[2] make $MgB_2$ an attractive candidate for practical applications. This has rejuvenated interest in the search for other similar materials, possibly with further enhanced properties. Some materials with related structure and otherwise similar properties have already been reported to superconduct[3,4,5], but all with lower transition temperatures.

Although $MgB_2$ may not fall neatly into either conventional or high temperature superconducting classes, it appears to be most similar to the conventional superconductors[6] but with a critical temperature far surpassing the 23 K $T_C$ of $Nb_3Ge$. Experiments have shown $MgB_2$ to exhibit a boron isotope effect [7,8], suggesting that phonons play an important role in electron-electron coupling. Specific heat measurements support phonon mediated coupling[9], but positive hall coefficients have been measured, more typical of high-$T_C$ superconductors[10].

Scanning tunneling spectroscopy (STS) is an excellent probe of the density of states in a superconductor. In particular, the differential conductance obtained through STS can determine the superconducting gap ($\Delta$) with a resolution much less than $k_BT$. Zero bias conductance peaks and other mid-gap states contain further information concerning the tunneling properties.

A number of tunneling spectroscopy measurements on $MgB_2$ have thus far been reported in the literature[11,12,13,14,15,16,17,18,19,20,21 ]. The results are varied and a consistent picture of the tunneling process on these materials has yet to be established. Reported values of the gap range from less than 2 meV to about 7.5 meV[11-19], well below and above the $\Delta=5.9$ meV value predicted from BCS weak coupling, $2\Delta/k_BT_C=3.52$. Most commonly reported gap values cluster around 2, 4, and 7 meV. In addition to variation in the magnitude, multiple gap features have been reported by some authors[12,18,19,20].

The existence of a double gap has been proposed to result from Cooper pairs associated with different sheets of the Fermi surface[22,23,24]. In this picture, two nearly-independent gap functions are weakly coupled such that they close at the same $T_C$. Of these two gaps, it is believed that one appears predominantly in the boron plane due to σ-bonded boron $p_x$ and $p_y$ orbitals, and the other results from the more three dimensional, π-bonded $p_z$ orbitals. A directional dependence in the tunneling then arises from the dimensionality of the orbitals associated with the particles in each band. Tunneling into the $p_z$ orbitals is nearly isotropic, whereas tunneling into the $p_x$ and $p_y$ orbitals is strongest for wavevectors in the boron plane. A recent report[20] suggests that spectral features are dependent on the directional configuration of the tunneling junction.

II. Experiment

We report STS measurements of the gap properties of ceramic $MgB_2$ and *c*-axis oriented epitaxial $MgB_2$ thin films in both S-I-S, formed in two ways, and normal metal-insulator-superconductor (N-I-S) configurations. Both types of S-I-S junction show a temperature-dependent zero-bias conductance peak and evidence of two superconducting gaps. One of the S-I-S type junctions resulted from inadvertently touching the normal metal Pt-Ir tip to the surface of a bulk ceramic $MgB_2$ sample. This led to tunneling between a piece of $MgB_2$ on the tip and the $MgB_2$ sample itself. We refer to this junction type as "type A" S-I-S junction. Ceramic samples were produced by direct reaction of Mg flakes with amorphous B powder, pressed into pellets, and fired (1 hour each at 600, 800, and 900° C) on a Ta foil[2].

The second type of S-I-S junction was prepared by affixing a piece of the ceramic sample to the Pt-Ir wire with Ag epoxy, thus forming a superconducting tip. This tip was then used for tunneling spectroscopy of an epitaxial, c-axis oriented $MgB_2$ thin film. This junction type will be referred to as "type B" S-I-S junction. We also performed N-I-S spectroscopy on the film using a Pt-Ir tip to independently determine its gap value. The thin film was grown by RF magnetron sputter deposition of amorphous B on (0001) $Al_2O_3$, followed by an anneal in Mg vapor at 850°C [25].

Tunneling characteristics were acquired using an Oxford Instruments CryoSXM in scanning tunneling microscopy (STM) mode from 5 K to 35 K. The sample and STM head are in the flowing He gas of a gas flow cryostat. The cryostat is mounted in an acoustically isolated enclosure supported by a vibration isolation table. Differential conductance (d$I$/d$V$) vs. bias voltage scans were obtained using a standard lock in technique. Voltage was quickly swept with the $z$-position feedback off in order to maintain a fixed position for the duration of each d$I$/d$V$. STM tips were mechanically prepared from 250 μm diameter Pt-Ir alloy wire, and also by attaching a piece of ceramic $MgB_2$ to the Pt-Ir wire.

Spectra were obtained at temperatures from 5 K to 30 K from the as-grown pellet for type A S-I-S junctions, and up to 35 K for type B S-I-S junctions. The N-I-S spectra of the film were obtained at $T$=5 K. In all cases, either altered surface structure or an impurity surface layer may have influenced the tunneling properties.

In the type A S-I-S junctions we were not able to maintain tunneling during some temperature changes, and found that the tip would occasionally touch the surface, potentially picking up a piece of the superconducting sample. This possibility requires us to discern the N-I-S or S-I-S nature of the junction from the d$I$/d$V$ measurement. We will show that these junctions are in fact S-I-S based on their temperature dependent zero bias conductance peaks and consistency with prepared type B S-I-S junction spectra at all temperatures. Some searching on the sample surface was necessary to obtain good tunneling conditions, obtained through topographic imaging and $I(V)$ sampling. The surface of our sample was found to be rough and only in some regions was consistent tunneling possible. In these regions of the sample we were able to achieve low noise tunneling with very consistent gap values and spectral features in general. In some regions we found Δ(T) to be approximately independent of position within experimental error, and in other regions the I(V) were consistently Ohmic.

As an aid in determining the junction type of the type A S-I-S junction we also performed S-I-S tunneling of type B S-I-S junctions consisting of an MgB$_2$ grain fabricated onto a tip tunneling into an MgB$_2$ thin film. We observed spectra consistent with the type A S-I-S junction at all temperatures including temperature dependent zero bias conductance peaks and similar gap values and spectral shapes with additional features for some spectra indicative of a second, larger gap (see section V). We attribute the presence of these features in only some but not all spectra to directional dependence of tunneling. We also performed N-I-S tunneling measurements on the film to independently determine its superconducting gap.

III. Data

Our d$I$/d$V$ curves for the type A S-I-S junctions exhibit a number of clear features and trends. We observe a nearly temperature-independent energy gap up to 30 K and a zero bias conductance peak (ZBCP) that increases with temperature. Most spectra have dips at voltages just greater than the gap energy and less tunneling than predicted by the BCS model for voltages less than 2Δ.

Each spectrum represents an average of typically 20 bias-voltage scans. The $z$-position feedback is off for the entire duration of the series of averaged scans (about 10 seconds). Our STM is especially vibration sensitive due to its very long piezoelectric scan tube, capable of attaining 8 μm scans at 4K. At low temperatures our data exhibit very sharp gap features, clearly identifiable gap features for higher temperatures, and ZBCPs that become more prominent with increasing temperature. Figure 1 shows an STS spectrum at 5 K with a fit using the BCS model appropriate for an S-I-S junction (Δ=2.2 meV, Γ=0.44 meV) (see Section V for details of the models). The gap is much smaller than the weak coupling limit.

Figure 2 shows an N-I-S spectrum of the MgB$_2$ film obtained with a Pt-Ir tip. This spectrum is fit with the N-I-S BCS model for the d$I$/d$V$ as described in section V with $\Delta_S$=2.3 meV and phenomenological smearing parameter $\Gamma$=0.8 meV. Note that the peaks occur at voltages slightly larger than Δ as obtained from the fit. This is due to broadening effects as is evidenced by the model. Gap values obtained from the peak position of sharply peaked gap features agree quite well with gap values obtained from fitting. Due to variable spectral shapes, including multiple gap features in some spectra, we were unable to fit all of the data. Therefore, average gap values we report here are obtained from peak locations. This leads to slightly inflated, but fairly representative values of the gap. These spectra are consistent with an N-I-S junction and in none of the spectra did we find evidence for picking up a piece of MgB$_2$ from the film surface.

The type B d$I$/d$V$ curves displayed characteristics similar to those for type A, and also, in some configurations, peaks at voltages consistent with $\Delta_S+\Delta_L$ from the two-band model. Here, $\Delta_S$ is the small gap associated with the π-band and $\Delta_L$ is the large gap associated with the σ-band. These different configurations are experimentally found at different locations on the sample, or at different tunneling distances. We argue that they arise from tunneling at different locations on the surface of the fabricated $MgB_2$ tip. Transitions between tunneling locations on the tip are not unusual in scanning tunneling microscopy, and can be attributed to imperfect topography of the tip. Figure 3 shows a d$I$/d$V$ characteristic with very similar shape as for type A. An S-I-S fit is also shown, and it gives a similar result for the gap, $\Delta_S$=2.4 meV. Figure 4 shows two d$I$/d$V$ characteristics with features indicative of two gaps. Although the spectra display dramatically different shapes, the gap features appear at nearly the same bias voltages in the two curves. We argue that the second gap feature observed should be associated with $\Delta_S+\Delta_L$, and its presence indicates that the piece of $MgB_2$ on the tip is oriented such that tunneling takes place through its *a-b* plane, as per the discussion in Section I. No peaks associated with $2\Delta_L$ were ever observed, consistent with the *c*-axis orientation of the film. The average gap values obtained from spectra of this type are $\Delta_S$=2.8 meV and $\Delta_L$=7.2 meV.

IV. General Observations

We measure an approximately constant superconducting gap for all tip locations and temperatures up to 30 K for both the ceramic sample and the epitaxial thin film. Figure 5 shows the mean gap measured at each temperature, indicating that there is little temperature dependence within our statistical error. However, the ZBCP changes dramatically with temperature and varies only slightly between different tip locations on the sample at the same temperature. Figure 6 shows that the magnitude of the ZBCP generally increases with increasing temperature. ZBCPs can result from a number of mechanisms, but most mechanisms lead to a ZBCP that increases in magnitude as the temperature decreases[26]. Based on the temperature dependence of our data and consistency with the type B data we will argue that the ZBCPs arise from an S-I-S tunneling configuration.

Deviations from BCS behavior for the dI/dV include a ZBCP (consistent with S-I-S but not N-I-S) and a dip and hump feature for voltages just greater than the gap. At all temperatures, some tunneling current is observed inside the gap. Although it may be reasonable at 5 K to attribute this to non-zero temperature, we find that this picture is inadequate at higher temperatures. In particular, Fig. 6 shows that the zero-bias conductance can be quite large at high temperatures. ZBCPs result from matching of peak density of states in the energy spectra between superconductors in the case of S-I-S tunneling, as discussed recently[21]. Note, also, the dip and hump feature for voltages of about 9 meV in Figs. 1 and 3. It has been suggested[21] that this feature is due to interband quasiparticle exchange in a two-band superconductor. This feature of the data may then be interpreted as evidence for a second gap.

V. Analysis

We have evaluated our type A S-I-S junction data using a BCS density of states, and two models for the tunneling characteristic, N-I-S and S-I-S. N-I-S and S-I-S fits are shown in Fig. 1. For all fits we use a modified BCS DOS, $N(E) = \text{Re} \frac{|E| - i\Gamma}{\sqrt{(E - i\Gamma)^2 - \Delta^2}}$, where $\Gamma$ is a smearing parameter. This is proportional to the N-I-S $dI/dV$ at zero temperature. Both the N-I-S and S-I-S models were fit with the expression, $\frac{dI}{dV} = G_0 \int \frac{d}{dV}(N(E)N(E+V)[f(E) - f(E+V)])dE$, where $N(E)$ is as given above, $f(E)$ is the Fermi function, and $G_0$ is the normal state conductance. For N-I-S tunneling, $N(E+V)$ is just the constant DOS for the normal metal at the Fermi level and is absorbed in the normalization. The S-I-S model is described by the entire expression with both DOS superconducting. This is formally where the two models differ.

Type A Junctions: Ceramic $MgB_2$-Insulator-Ceramic $MgB_2$

For our type A junction data we find that the N-I-S model (see Fig. 1) fits satisfactorily with $\Delta$=3.9 meV (approximately twice the gap value of our later S-I-S interpretation), where we have used the ambient measurement temperature for $T$ in the Fermi function (see Fig. 1). However, this model does not account for the ZBCP detected at all temperatures above 5 K (see Fig. 6). If the junction is N-I-S, then we conclude that the model only describes part of the characteristic and we must consider deviations from the ideal case described by the model.

In light of the possibility of having a piece of $MgB_2$ on the STM tip we also fit our data with the BCS model for an S-I-S junction. Our 5 K data were fit using $T$=5 K in the Fermi function and smearing parameter $\Gamma$=0.44 meV, determined by manually optimizing the numerical integration of the model $dI/dV$. This picture is more consistent with our data and qualitatively accounts for the temperature dependence of the ZBCP. This can be seen by considering the semiconductor model of S-I-S tunneling. In this model, peaks in the $dI/dV$ will occur at $\pm(\Delta_1+\Delta_2)$ and $\pm(\Delta_1-\Delta_2)$, with $\Delta_1$ and $\Delta_2$ the gap values of the potentially dissimilar superconducting electrodes. In the case of two superconductors of the same material, $\Delta_1=\Delta_2\equiv\Delta$, eV=$\pm(\Delta_1-\Delta_2)$=0 corresponds to the ZBCP. The temperature dependence of the ZBCP results from thermally excited quasiparticles in one superconducting electrode tunneling into empty states above the gap of the other superconductor. At zero temperature, the complete absence of quasiparticles suppresses the ZBCP. Higher temperatures provide the thermal energy necessary to excite more quasiparticles, increasing the ZBCP. The peak in the dI/dV occurs at zero bias due to the peaked nature of the DOS at the gap edge. Peaks at $\pm(\Delta_1+\Delta_2)=2\Delta$ results from Cooper pair breaking with one quasiparticle tunneling immediately while the other is promoted to lower lying quasiparticle states in the same superconductor.

Type B Junctions: Ceramic $MgB_2$-Insulator-Epitaxial $MgB_2$ Thin Film

Figure 7 shows the mean gap measured at each temperature up to 35 K for type B junctions. Again, the gap is approximately constant up to 30 K, consistent with our observations for type A junctions. The value of the gap is also consistent with the S–I–S interpretation of the type A junctions. Figure 8 displays offset d$I$/d$V$ curves for this junction type at all measured temperatures. The spectral shape, including ZBCPs and the dip and hump feature, as well as peak heights and locations closely resembles those for type A. We take this consistency as strong evidence for the S-I-S tunneling interpretation of the type A junctions on bulk ceramic material.

The semiconductor model of S-I-S tunneling suggests that peaks should occur at the sums and differences of the gaps. Multiple gap features in some of our data are consistent with this two gap picture. In the context of a double gap superconductor we would expect peaks in the d$I$/d$V$ for voltages of $\pm 2\Delta_S$, $\pm(\Delta_S\pm\Delta_L)$, $\pm 2\Delta_L$, and $\pm(\Delta_S-\Delta_S)=\pm(\Delta_L-\Delta_L)=0$. We, however, have not observed $\pm 2\Delta_L$ and only in some spectra do we observe $\pm(\Delta_S+\Delta_L)$. It has been suggested that, due to tunneling matrix elements, the small gap is more geometrically favorable to appear from a random junction orientation than is the large gap[20]. Spectra exhibiting only the peaks at $\pm 2\Delta_S$ are consistent with predominantly c-axis tunneling and those spectra also exhibiting peaks at $\pm(\Delta_S+\Delta_L)$ are consistent with an $a-b$ plane component from the tunneling tip. From the geometric, matrix element considerations presented in section I it is then clear that we should, in fact, *not* observe $\pm 2\Delta_L$ due to the c-axis orientation of the film. As we have discussed above, the ZBCP is temperature dependent and thus its presence or absence in any spectrum is determined by temperature. That leaves $\pm(\Delta_S-\Delta_L)$, which happens to be approximately equal to $\pm 2\Delta_S$. We argue that the $\pm(\Delta_S-\Delta_L)$ peak is obscured by the stronger $\pm 2\Delta_S$ peak, though some broadening will result, since $\Delta_S^{bulk}$ is not exactly equal to $\Delta_S^{film}$. The $T_C$ of the film (35 K) is slightly depressed from the bulk value (39 K), which means that the gap of the film should also be slightly depressed from the bulk value. By comparing Figs. 1 and 3 we notice that some additional broadening occurs when tunneling takes place between bulk and film.

VI. Conclusions

In summary, we have measured the superconducting gap of bulk and thin film $MgB_2$ by low temperature scanning tunneling spectroscopy in both S-I-S and N-I-S configurations. The gap determined from an overall average of our 5 K measurements is $\Delta=2.3$ meV, smaller than the BCS weak coupling limit, and varies little with temperature up to $T/T_C=0.77$. We have also observed evidence for a second, larger gap ($\Delta_L=7.2$ meV) for prepared S-I-S junctions. The simple semiconductor picture extended to the two-band superconductor model succesfully describes the peak locations in our data. The detailed shape of the spectra, including the dip and hump feature observed just above the gap features, are not well-modeled by any of the fits investigated here, and may arise through microscopic mechanisms not considered here. The most ubiquitous 'non-gap' feature we have observed, the dip and hump structure, has recently been attributed to a form of interband interaction[21]. We do not have an understanding of the other more subtle shape differences between spectra. However we do observe that gap features are robustly independent of the detailed spectral shape, and give a consistent representation of the gap structure in $MgB_2$.

Work at the University of Wisconsin was supported by the NSF MRSEC on Nanostructured Materials.

FIGURE CAPTIONS

Figure 1. Differential conductance of ceramic MgB$_2$ at 5 K for the type A S-I-S junction (data are closed circles) with fits to the BCS S-I-S junction (solid line; $\Delta$=2.2 meV, $\Gamma$=0.44 meV) and N-I-S junctions (dashed line; $\Delta$=3.9 meV, $\Gamma$=0 meV) models.

Figure 2. Differential conductance of MgB$_2$ thin film (data are closed circles) with BCS N-I-S fit with $\Delta$=2.3 meV and smearing $\Gamma$=0.8 meV. (fit is solid line).

Figure 3. Differential conductance of ceramic MgB$_2$ at 5 K for the type B S-I-S junction (data are closed circles) with a BCS S-I-S fit with $\Delta$=2.75 meV and $\Gamma$=0.8 meV. Note the striking similarity with Figure 1.

Figure 4. Two overlaid differential conductance curves of ceramic MgB$_2$ at 5 K for the type B S–I–S junction. These data display the consistency with which multiple gap features appear at the same voltages despite different peak heights. The dashed lines are guides to the eye.

Figure 5. Temperature dependence of the superconducting gap for the type A S-I-S junction averaged for all data at each temperature. Error bars are one sigma.

Figure 6. Representative d$I$/d$V$ curves of the type B S-I-S junction for the as grown sample for temperatures ranging from 5 K to 25 K (offset for clarity (2 units for 10 K curve and 1 unit for each additional temperature change as indicated by offset zero lines)). Note the general trend of increasing ZBCP height with increasing temperature and the relatively constant gap width.

Figure 7. Temperature dependence of the superconducting gap for the type B S-I-S junction averaged for all data at each temperature. Error bars are one sigma.

Figure 8. Representative d$I$/d$V$ curves of the type B S-I-S junction for the MgB$_2$ thin film for temperatures ranging from 5 K to 35 K (offset for clarity (2 units for 10 K curve and 1 unit for each additional temperature change as indicated by offset zero lines )). Note the striking similarity to Fig. 6 in the general trend of increasing ZBCP height with increasing temperature and the relatively constant gap width.

Figure 1
Heitmann et al
Superconductor Science and Technology

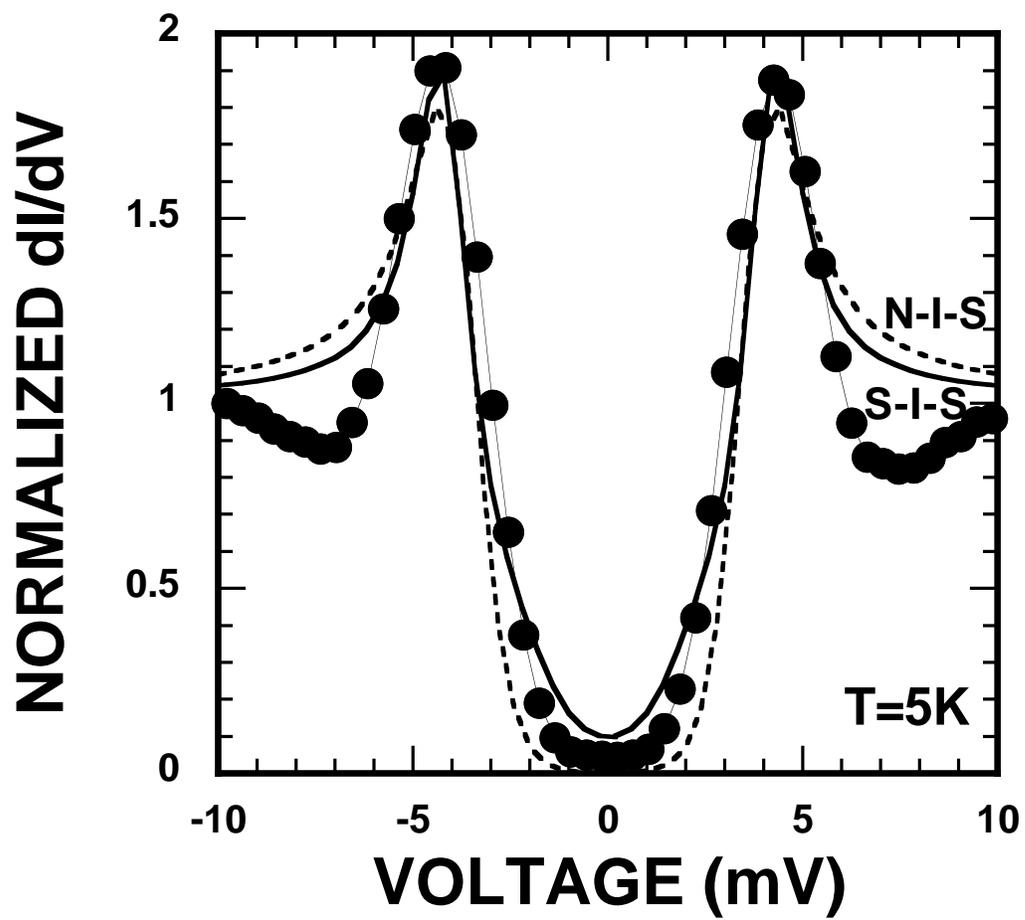

Figure 2
Heitmann et al
Superconductor Science and Technology

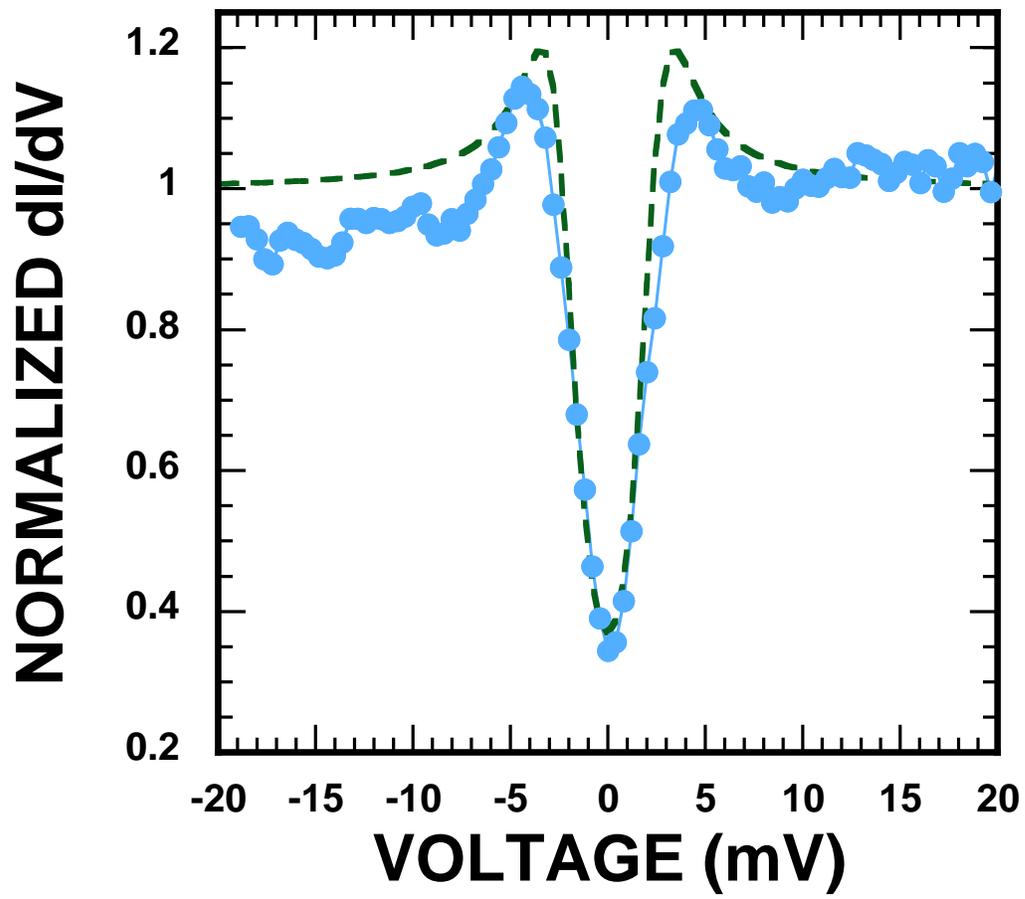

Figure 3
Heitmann et al
Superconductor Science and Technology

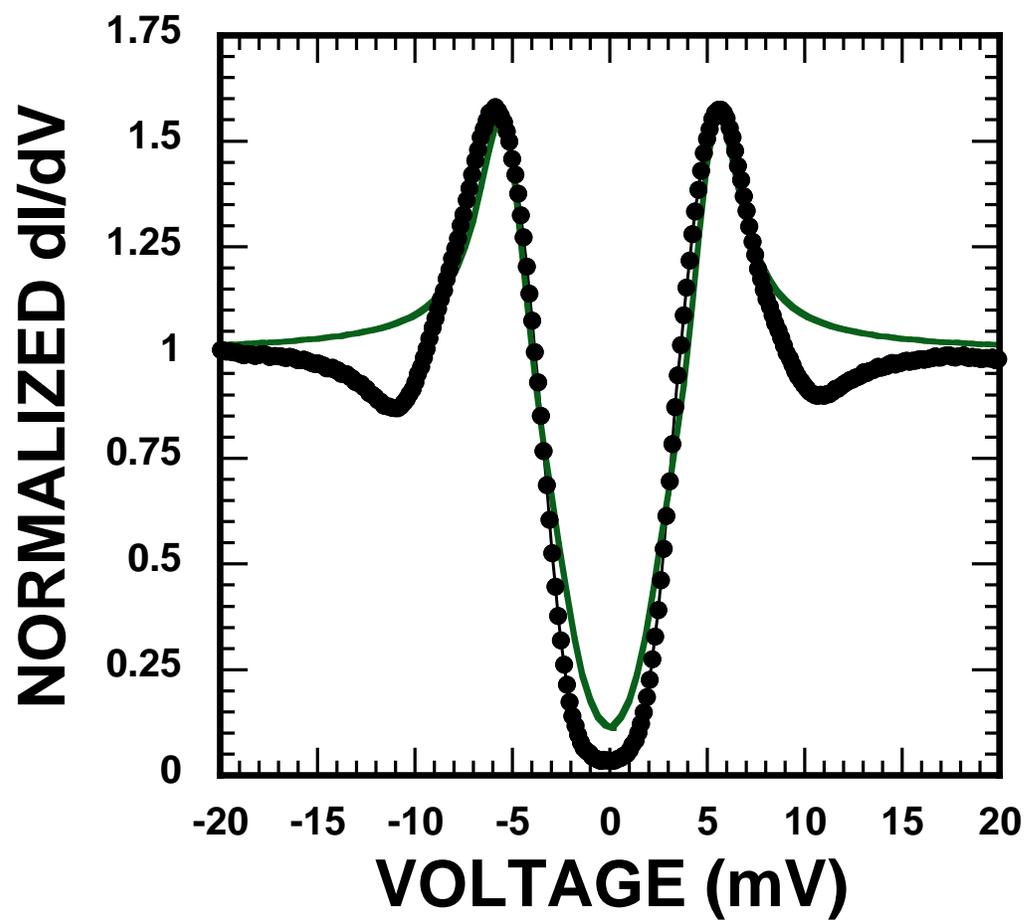

Figure 4
Heitmann et al
Superconductor Science and Technology

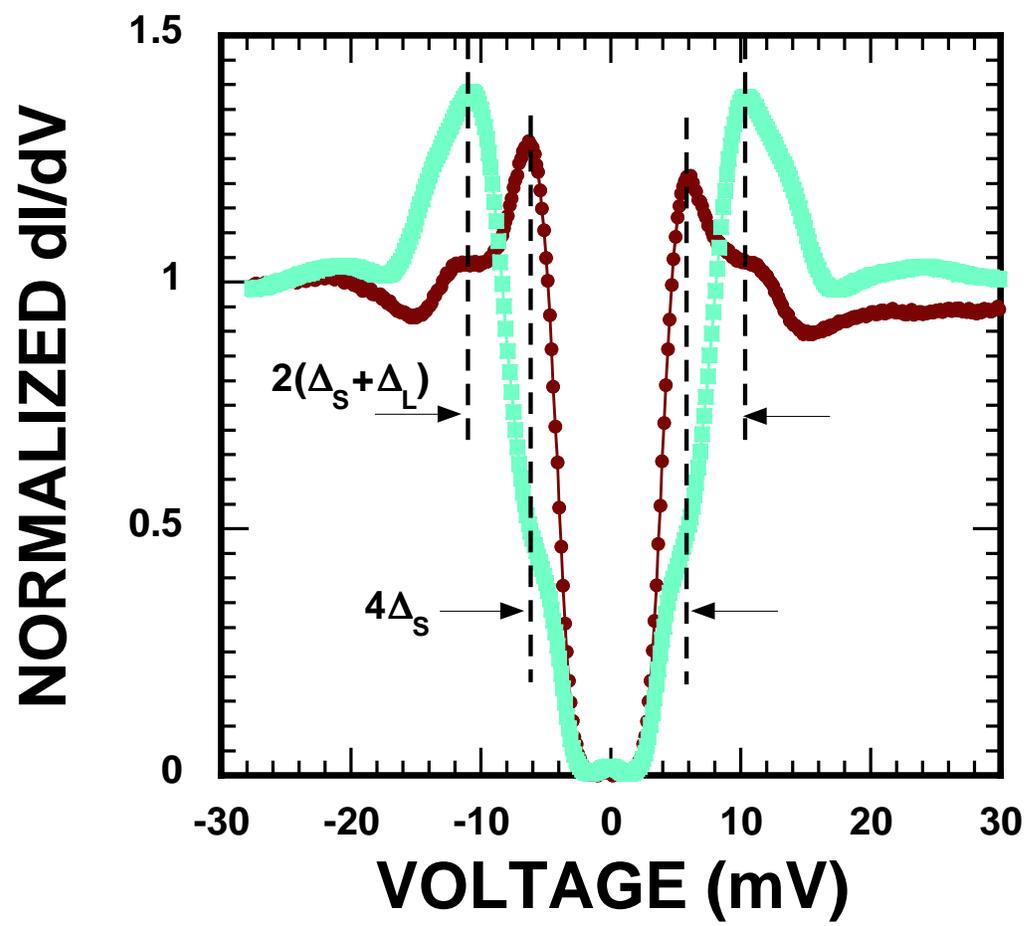

Figure 5
Heitmann et al
Superconductor Science and Technology

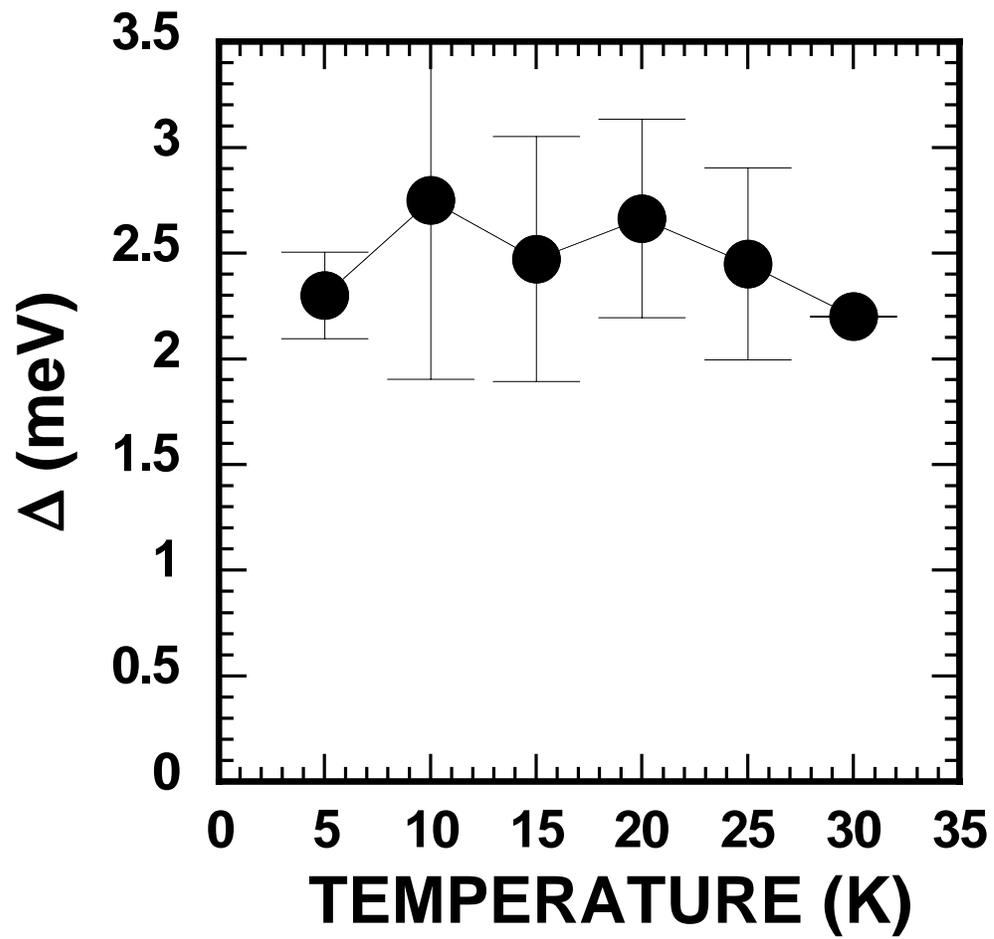

Figure 6
Heitmann et al
Superconductor Science and Technology

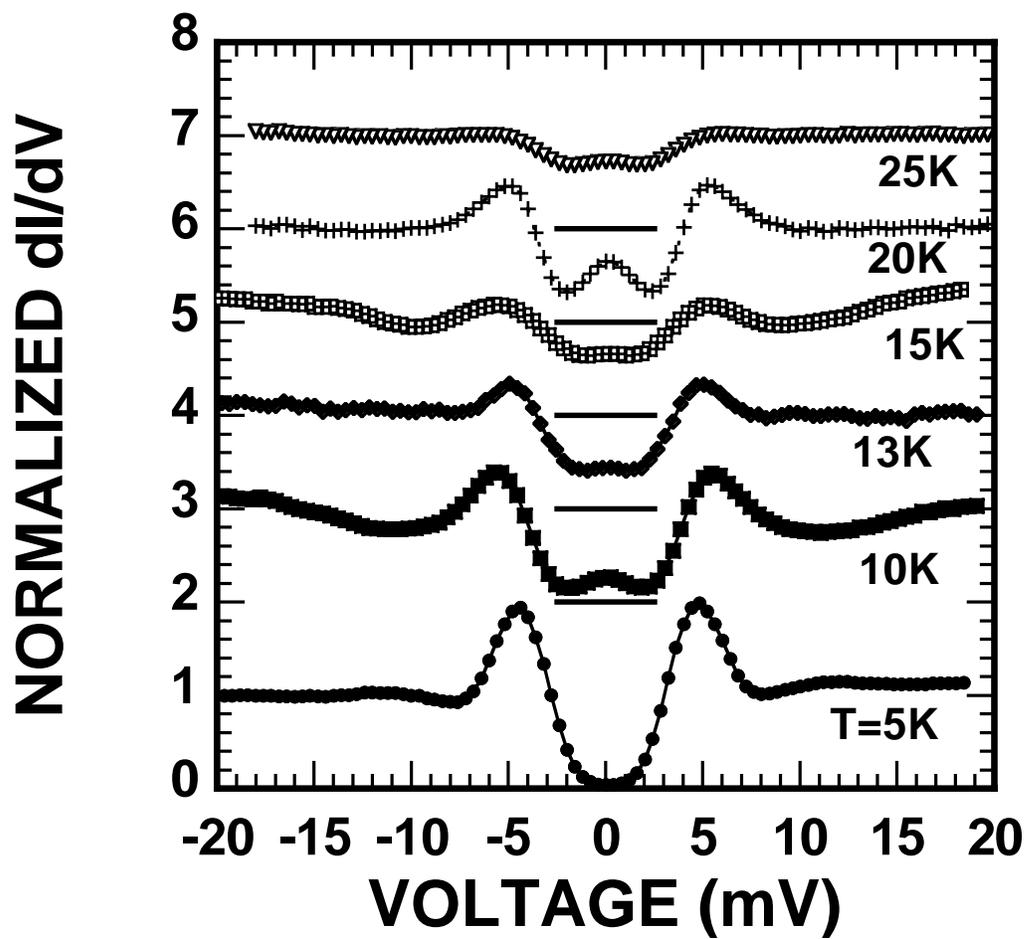

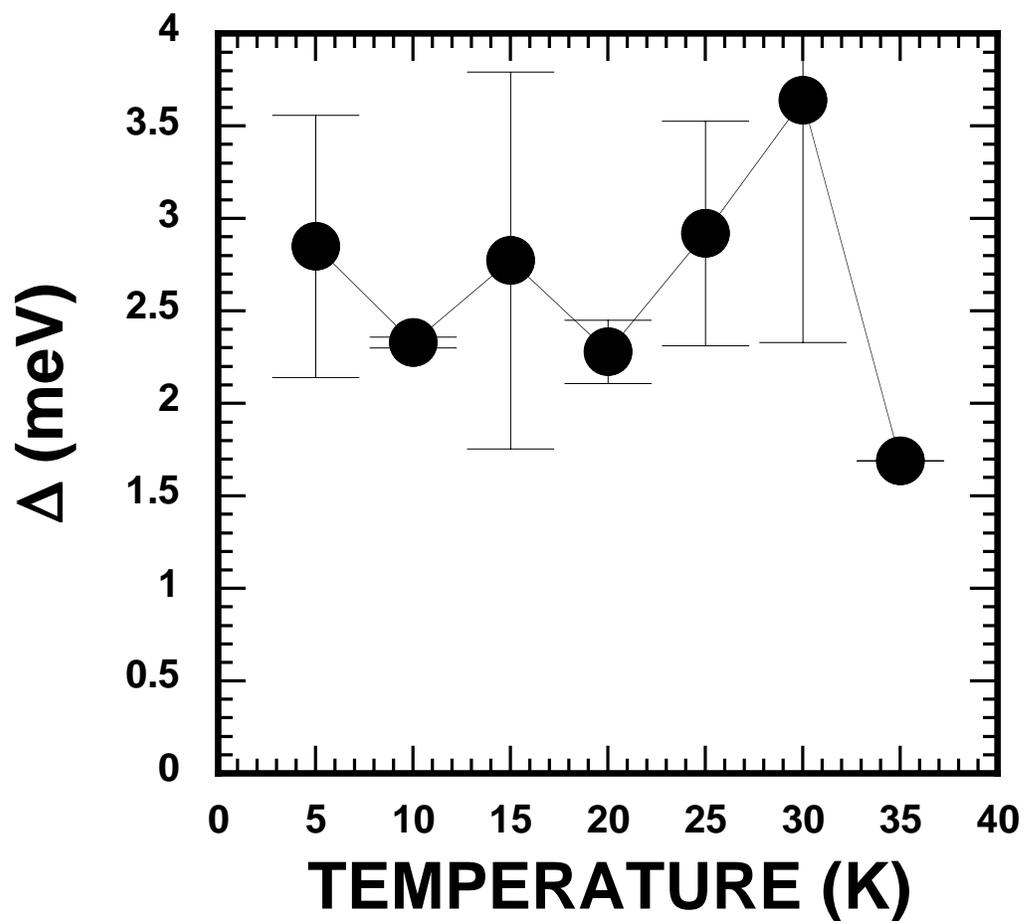

Figure 8
Heitmann et al
Superconductor Science and Technology

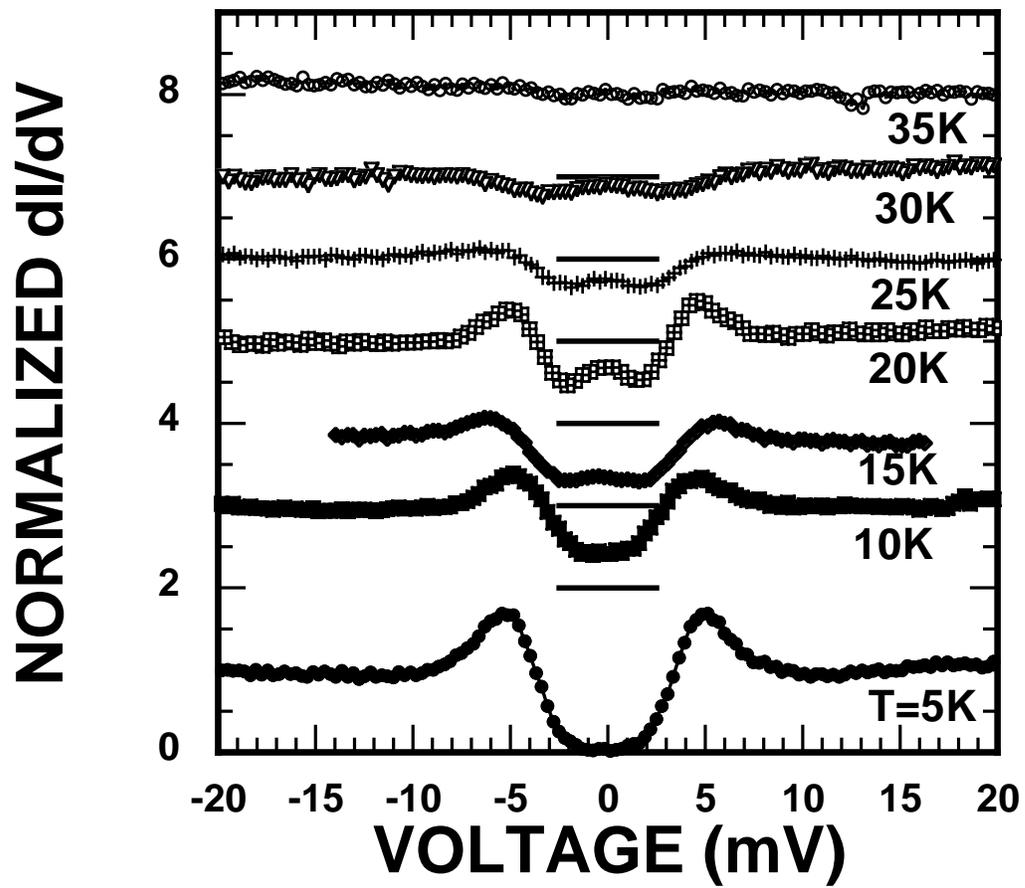